\documentstyle[12pt]{article}

\newcommand{\e}{\varepsilon}
\newcommand{\p}{\bot}
\newcommand{\dd}{\partial}
\newcommand{\de}{\delta}
\newcommand{\ls}{\left(}
\newcommand{\lks}{\left[}
\newcommand{\rs}{\right)}
\newcommand{\rks}{\right]}
\newcommand{\ra}{\rangle}
\newcommand{\str}[1]{\mathrel{\mathop{\longrightarrow}\limits_{#1}}}

\newcommand{\disn}[2]{$$\displaylines{\refstepcounter{equation}%
            \label{#1}\hskip 1em minus 1em #2\hfilneg}$$}
\newcommand{\nom}{\hfil\hskip 1em minus 1em (\theequation)}
\newcommand{\no}{\hfil \hskip 1em minus 1em\phantom{(\theequation)}%
            \hfilneg\cr\hfilneg\hskip 1em minus 1em\hfil}

\makeatletter

\newcount\dobav
\newcount\otlog
\newcount\vrem
\def\@citex[#1]#2{\if@filesw\immediate\write\@auxout{\string\citation{#2}}\fi
  \let\@citea\@empty
  \dobav=-1
  \otlog=-1
  \@cite{\@for\@citeb:=#2\do
    {\def\@tempa##1##2\@nil{\edef\@citeb{\if##1\space##2\else##1##2\fi}}%
     \expandafter\@tempa\@citeb\@nil
     \@ifundefined{b@\@citeb}{\@warning%
       {Citation `\@citeb' on page \thepage \space undefined}%
       \vrem=-1}{\vrem=\csname b@\@citeb\endcsname}
\advance\vrem by -1 \ifnum \vrem=\dobav
 \otlog=\vrem
 \advance\otlog by 1
\else
 \ifnum \vrem=\otlog
  \advance\otlog by 1
 \else
  \ifnum \otlog>0
   \advance\dobav by 1
   \ifnum \otlog=\dobav
    \hbox{,\penalty\@m\ \the\otlog}%
   \else
    \hbox{--\the\otlog}%
   \fi
   \otlog=-1
  \fi
  \dobav=\vrem
  \advance\dobav by 1
  \@citea\def\@citea{,\penalty\@m\ }%
  \ifnum \dobav=-1
   {\reset@font\bf ?}%
  \else
   \hbox{\the\dobav}%
  \fi
 \fi
\fi
}%
\ifnum \otlog>0
 \advance\dobav by 1
 \ifnum \otlog=\dobav
  \hbox{,\penalty\@m\ \the\otlog}%
 \else
  \hbox{--\the\otlog}%
 \fi
\fi }{#1}}

\long\def\@makecaption#1#2{%
   \vskip 10\p@
   \setbox\@tempboxa\hbox{#1. #2}%
   \ifdim \wd\@tempboxa >\hsize
       #1. #2\par
     \else
       \hbox to\hsize{\hfil\box\@tempboxa\hfil}%
   \fi}

\renewcommand{\section}{\@startsection{section}{1}{0pt}%
          {3.5ex plus 1ex minus .2ex}{2.3ex plus .2ex}{\noindent\hfil\bf}}

\makeatother

\begin{document}
\large

\title{Gauge Invariant Regularization\\
of Quantum Field Theory on the Light-Front\\}

\author{S.~A.~Paston\thanks{E-mail: paston@pobox.spbu.ru},
E.~V.~Prokhvatilov\thanks{E-mail: Evgeni.Prokhvat@pobox.spbu.ru},
V.~A.~Franke\thanks{E-mail: franke@pobox.spbu.ru}\\
St.-Petersburg State University, Russia}

\date{\vskip 15mm}

\maketitle

\begin{abstract}
\normalsize
Gauge invariant regularization of quantum field theory in the
framework of Light-Front (LF) Hamiltonian formalism via
introducing a lattice in transverse coordinates and imposing
boundary conditions in LF coordinate $x^-$ for gauge fields on
the interval $|x^-|\le L$ is considered. The remaining
ultraviolet divergences in the longitudinal momentum $p_-$ are
removed by gauge invariant finite mode regularization. We find
that LF canonical formalism for the introduced regularization
does not contain usual most complicated second class constraints
connecting zero and nonzero modes of gauge fields. The described
scheme can be used either for the regularization of conventional
gauge theory or for gauge invariant formulation of effective
low-energy models on the LF. The lack of explicit Lorentz
invariance in our approach leads to difficulty with defining the
vacuum state. We discuss this difficulty, particulary, in the
connection with the problem of taking the limit of continuous
space.
\end{abstract}

\newpage
\section{Introduction}
Canonical formulation of field theory in
Light-Front (LF) coordinates
$x^{\pm}=(x^0\pm x^3)/\sqrt{2}, x^1, x^2$,
where $x^0, x^1, x^2, x^3$ are Lorentz coordinates, was proposed by Dirac
\cite{dirak}.
The $x^+$ plays the
role of time, and canonical quantization is carried out on a hypersurface
$x^+=const$.
The advantage of this scheme is connected with the positivity of the
momentum $P_-$ (translation operator along $x^-$ axis), which becomes
quadratic in fields on the LF. As a consequence the lowest eigenstate
of the operator $P_-$ is both physical vacuum and the "mathematical" vacuum
of perturbation theory. Using Fock space over this vacuum one can
solve
stationary Schroedinger equation with Hamiltonian $P_+$
(translation operator
along $x^+$ axis) to find the spectrum of bound states. The problem of
describing the physical vacuum, very complicated in usual formulation
with Lorentz coordinates, does not appear here. Such approach, based on
solving Schroedinger equation on the LF, is called LF Hamiltonian approach.
It attracts attention for a long time as a possible mean for solving
Quantum Field Theory nonperturbatively.

  While giving essential advantages, the application of LF coordinates in
Quantum Field Theory  leads to some difficulties. The hyperplane
$x^+=const$
 is a characteristic  surface for relativistic differential field equations.
It is not evident without additional investigation that the
quantization on such a hypersurface generates a theory equivalent to one
quantized in usual way in Lorentz coordinates
\cite{chang1,lenc,pred1,pred2,pirner,bur1,bur2}.
This is essential, in particular,
because of special divergences at $p_-=0$ appearing in LF
quantization scheme. Beside of conventional
ultraviolet regularization one has to
apply special regularization of these divergences. Usually the
following simplest prescriptions of such regularization are considered:

(a) cutoff of the momenta $p_-$
 \disn{1}{
 |p_- |\ge\e ,\quad\e>0;
 \nom}

(b) cutoff of the coordinate $x^-$
 \disn{2}{
 -L\le x^-\le L
 \nom}
with periodic boundary conditions in $x^-$ for all fields.

The prescriptions (a) and (b)   are
convenient for Hamiltonian approach,
but both of them break Lorentz
invariance, and the prescription (a) breaks  also the gauge invariance.

The problem of constructing a LF Hamiltonian which generates a theory
equivalent to original  Lorentz and gauge invariant one turned out to be
rather difficult. Nevertheless it can be solved
perturbatively to all orders for nongauge field
theories \cite{tmf97,nucl02} and also
for gauge theories (including QCD) \cite{tmf99,qedpv,qedhep,tmf02}
via gauge-noninvariant
regularization procedure (under the prescription (a))
using special methods
to restore the symmetries in the limit of removed regularization.

The regularization prescription (b) discretizes the spectrum of
the operator $P_-$ ($p_-=\pi n/L$, where $n$ is an integer).
This formulation
is called sometimes "Discretized Light Cone
Quantization  (DLCQ)" \cite{brodsk}.
Fourier components of
fields, corresponding to $p_-=0$
(and usually called "zero modes")  turn out
to be dependent variables and are to be expressed in terms
of nonzero modes  via
solving constraint equations (constraints) \cite{nov2,nov2a,nov3,nov1}.
These constraints are usually very
complicated due to their nonlinearity in fields,
and solving of them is a difficult problem. Moreover, in quantum theory one
encounters the uncertainty in ordering of operators in the expressions
for these constraints.

The introduction of space-time lattice for gauge-invariant regularization
of nonabelian gauge theories is well known \cite{wils}.
Gauge invariant regularization in continuous space-time is also known
\cite{halp} but it seems not suitable for the LF quantization.
For the LF formulation only
the lattice in transverse coordinates $x^1, x^2$ is used.
In this formulation
it is convenient to define variables so as to have the action
polynomial in these variables \cite{barpir1,barpir2,heplat}.
Such a regularization is not Lorentz
invariant, and one can only hope that the Lorentz invariance
can be restored in continuous space limit.
Nevertheless many attempts to apply LF Hamiltonian formulation with the
prescription (b), combined with transverse space
lattice, are undertaken (for "color dielectric" type models
\cite{dalley1,dalley4,pirner1,mack}).
In all of these works zero modes of fields are thrown out,
so that, in fact, gauge invariance is violated.

In the present paper we consider canonical  LF formulation
of gauge theories,
regularized in gauge-invariant way. To achieve this goal we
introduce transverse space
lattice, discretize the momentum $p_-$ according to the prescription
(b) (with all zero modes of fields included) and apply the so called
"finite mode" ultraviolet
regularization in $p_-$. The last means a cutoff in eigen values of
covariant derivative operator $D_-$  in the expansion of lattice field
variables in eigen functions of this operator. These field
variables are  lattice modification of transverse components of usual gauge
fields. They are described by complex matrices, defined on lattice links.
Only such variables
admit mentioned above "finite mode" regularization (for fermion fields
analogous method was applied in \cite{konmod1,konmod2}).

It is interesting that in the framework of this formulation one can avoid
complicated canonical 2nd-class constraints, usually present in canonical
LF formalism in continuous space. This greatly simplifies canonical
quantization
procedure. However the absence of explicit Lorentz invariance of
the regularization scheme makes the investigation  of the connection with
conventional Lorentz-covariant  formulation difficult. In particular,
there is a problem of the description of quantum vacuum  as common lowest
eigen state of both operators $P_-$ and $P_+$.  This question is discussed
at the end of this paper.

\section{Gauge-invariant action on the transverse lattice}

At first we introduce particular ultraviolet regularization via
a lattice in transverse coordinates $x^1, x^2$ and choose variables
so as to have the action, which is polynomial in these variables
\cite{barpir1,barpir2}. Furthermore, we use the described
gauge-invariant regularization (b) of
singularities at $p_- \to 0$  and gauge-invariant ultraviolet
cutoff in modes of covariant derivative $D_-$
(then ultraviolet regularization of the
theory is complete). For simplicity we consider  $U(N)$ theory of
pure gauge fields although the generalization to
$SU(N)$-theory or a theory with fermions is not difficult.

The components of gauge field along continuous coordinates $x^+$, $x^-$
can be taken without a modification and related to the sites of the
lattice. Transverse components are described by complex $N\times N$
matrices $M_k(x)$,
$k=1,2$. Each matrix $M_k(x)$ is related to the link
directed from the site $x-e_k$ to the site $x$.
The transverse vector $e_k$ connects two neighbouring sites on the lattice
being directed along the positive axis $x^k$ ($|e_k|\equiv a$),
see fig.~1.
 \begin{figure}[ht]
\vskip 8mm
\input FIG1.PIC
\caption{}
\end{figure}
The matrix $M_k^+(x)$ is related to the same link but
with opposite direction, see fig.~2.
 \begin{figure}[ht]
\vskip 8mm
\input FIG2.PIC
\caption{}
\end{figure}
In the following for the index $k$ the usual rule
of summation on repeated indices is not used, and, where it is necessary,
the symbol of a sum is indicated.

The elements of these matrices are considered as independent variables.
This makes the action polynomial.
For any closed directed loop on the lattice we can construct the trace of
the product of  matrices $M_k(x)$ sitting on the links
and order from the right to the left along this loop. For
example the expression
 \disn{3}{
 {\rm Tr\;}\left\{M_2(x)M_1(x-e_{2})M_2^+(x-e_{1})M_1^+(x)\right\}
 \nom}
is related to the loop shown in fig.~3.
 \begin{figure}[ht]
\vskip 8mm
\input FIG3.PIC
\caption{}
\end{figure}

It should be noticed that a product of matrices
related to closed loop, consisting of one and the
same link passed in both directions, is not identically unity because the
matrices $M_k$ are not unitary (see, for example, fig.~4).
 \begin{figure}[ht]
\vskip 8mm
\input FIG4.PIC
\caption{}
\end{figure}

The unitary matrices $U(x)$ of gauge transformations act on the $M$
and $M^+$ in the following way:
 \disn{5.1a}{
M_k(x)\to M'_k(x)=U(x)M_k(x)U^+(x-e_{k}),
\nom}
 \disn{1b}{
M_k^+(x)\to M'^+_k(x)=U(x-e_{k})M_k^+(x)U^+(x).
\nom}
A trace of the product of the matrices, related to a closed loop along
lattice links, is invariant with respect to
these transformations.
To relate the matrices $M_k$ with usual gauge fields of continuum theory
let us write these matrices in the following form:
 \disn{5.2a}{
M_k(x)=I+gaB_k(x)+igaA_k(x),\qquad B_k^+=B_k,\quad A_k^+=A_k.
\nom}

Then in the $a\to 0$ limit the fields $A_k(x)$ coincide with transverse gauge
field components, and the $B_k(x)$  turn out to be extra (nonphysical) fields
which should be switched off in the limit. Below we show how to get this.

The analog of the field strength
 \disn{5.2b}{
 F_{\mu\nu}=\dd_{\mu}A_{\nu}-\dd_{\nu}A_{\mu}-ig[A_{\mu},A_{\nu}],
 \nom}
multiplied by $i$, can be defined as follows:
 \disn{5.n1}{
G_{+-}=iF_{+-},\qquad F_{+-}(x)=\dd_+A_-(x)-\dd_-A_+(x)-ig[A_+(x),A_-(x)],\no
G_{\pm,k}(x)=\frac{1}{ga}\lks \dd_{\pm}M_k(x)-ig\ls A_{\pm}(x)M_k(x)-
M_k(x)A_{\pm}(x-e_k)\rs\rks,\no
G_{12}(x)=-\frac{1}{ga^2}\lks M_1(x)M_2(x-e_1)-M_2(x)M_1(x-e_2)\rks.
\nom}
 Under gauge transformation these quantities transform as follows:
 \disn{5.n2}{
G_{+-}(x)\to G'_{+-}(x)=U(x)G_{+-}(x)U^+(x),\no
G_{\pm,k}(x)\to G'_{\pm,k}(x)=U(x)G_{\pm,k}(x)U^+(x-e_k),\no
G_{12}(x)\to G'_{12}(x)=U(x)G_{12}(x)U^+(x-e_1-e_2).
\nom}

We choose a simplest form of the action having correct naive continuum
limit:
 \disn{5.n3}{
S=a^2\sum_{x^\p}\int\! dx^+\!\int\limits^L_{-L}\! dx^-\;{\rm Tr}
\lks G^+_{+-}G_{+-}+\sum_k\ls G^+_{+k}G_{-k}+
G^+_{-k}G_{+k}\rs -G^+_{12}G_{12}\rks+\hfill\no\hfill+S_m,
\nom}
where the additional term $S_m$ gives an infinite mass to extra fields $B_k$
in $a\to 0$ limit:
 \disn{5.n4}{
S_m=-\frac{m^2(a)}{4g^2}\sum_{x_\p}\int dx^+
\int\limits^L_{-L}dx^-\sum_k{\rm Tr}\lks\ls M_k^+(x)M_k(x)-I\rs^2\rks
\str{a\to 0}\hfill\no\hfill
\str{a\to 0} -m^2(a)\int d^2x^\p\int dx^+
\int\limits^L_{-L}dx^-\sum_k{\rm Tr} \ls B^2_k\rs,\qquad m(a)\str{a\to 0}\infty.
\nom}
It is supposed that this leads to necessary decoupling of the fields $B_k$.

\section{Canonical quantization on the Light Front}

   Let us fix the gauge as follows:
 \disn{5.n5}{
\dd_-A_-=0,\qquad A_-^{ij}(x)=\de^{ij}v^j(x^\p,x^+).
\nom}
For simplicity below we denote the argument of quantities,
not depending on the $x^-$, again by  $x$.
Let us remark that starting with arbitrary field $A_\mu$,
periodic in $x^-$, it is not possible to take
zero modes of the $A_-$ equal to
zero without a violation of the periodicity.
But it is possible to make the $A_-$
diagonal as in the eq.-n (\ref{5.n5}) \cite{nov2,nov2a,nov3,nov1}.

Then the action (\ref{5.n3}) can be written in the  form:
 \disn{5.n6}{
S=a^2\sum_{x^\p}\int dx^+\int\limits^L_{-L}dx^-\left\{
\sum_i\lks 2F^{ii}_{+-}(x)\dd_+ v^i(x)\rks+\right.\hfill\no
+\frac{1}{(ga)^2}\sum_{i,j}\sum_k\lks D_-{M_k^{ij}}^+(x)\dd_+M_k^{ij}(x)
+h.c.\rks+\no\hfill
+\left.\sum_{i,j} A_+^{ij}(x)Q^{ji}(x)-{\cal H}(x)\right\},
\nom}
where
 \disn{5.n7}{
D_-M_k^{ij}(x)\equiv \ls \dd_--igv^i(x)+igv^j(x-e_k)\rs M^{ij}_k(x),\no
D_-{M_k^{ij}}^+(x)\equiv \ls \dd_-+igv^i(x)-igv^j(x-e_k)\rs {M^{ij}_k}^+(x),\no
D_-F_{+-}^{ij}(x)\equiv \ls \dd_--igv^i(x)+igv^j(x)\rs F^{ij}_{+-}(x),
\nom}
$A_+^{ij}(x)$ play the role of Lagrange multipliers,
 \disn{5.n8}{
Q^{ji}(x)\equiv 2D_-F_{+-}^{ji}(x)+\hfill\no\hfill
+\frac{i}{ga^2}\sum_{j'}\sum_k
  \lks {M^{ij'}_k}^+(x) D_-M_k^{jj'}(x)-
{M_k^{j'j}}^+(x+e_k)D_-M_k^{j'i}(x+e_k)\rks=0,
\nom}
are gauge constraints and
 \disn{5.n81}{
{\cal H}=\sum_{ij}\ls {F_{+-}^{ij}}^+F_{+-}^{ij} +
{G_{12}^{ij}}^+G_{12}^{ij}\rs+{\cal H}_m
\nom}
is the Hamiltonian density. The term ${\cal H}_m$ can be obtained from the
expression (\ref{5.n4}) in standard way.

The constraints  can be resolved explicitly by expressing the $F_{+-}^{ij}$
in terms of other variables, but
zero mode components $F^{ii}_{+-(0)}$
 can not be found from constraint equations and play the role of independent
canonical variables. Zero modes
$Q^{ii}_{(0)}(x^{\p},x^+)$ of the constraints
remain unresolved and
 are imposed as  conditions on physical states:
 \disn{5.4.9}{
Q^{ii}_{(0)}(x^\p,x^+)\left|\Psi_{phys}\right>=0.
\nom}
   In order to find a set of independent canonical variables we write
Fourier transformation in $x^-$ of the fields $M_k^{ij}(x)$ in the following
form:
 \disn{5.4.10}{
M^{ij}_k(x)=\frac{g}{\sqrt{4L}}\sum_{n=-\infty }^{\infty }
\left\{ \Theta \ls p_n+gv^i(x)-
gv^j(x-e_k)\rs M^{ij}_{nk}(x^\p,x^+)+\right. \hfill\no
\left. +\Theta \ls -p_n-gv^i(x)+
gv^j(x-e_k)\rs {M^{ij}_{nk}}^+(x^\p,x^+) \right \} \times \no\hfill
\times \left | p_n+gv^i(x)-
gv^j(x-e_k) \right |^{-1/2} e^{-ip_nx^-},
\nom}
where
 \disn{5.n9}{
\Theta (p) = \cases{
$1$, & $p > 0$ \cr
$0$, & $p < 0$ \cr},
\qquad p_n=\frac{\pi}{L}\,n, \quad n\epsilon Z.
\nom}
Due to the gauge (\ref{5.n5})
this Fourier transformation coincides with the expansion in
eigen modes of the operator $D_-$. Therefore the ultraviolet cutoff in
these modes, which we shall apply, reduces to the following condition on the
number of terms in the sum (\ref{5.4.10}):
 \disn{5.15a}{
 |p_n + gv^i(x) - gv^j(x-e_k) | \le \frac{\pi}{L}\,\bar n,
 \nom}
where the $\bar n$ is integer parameter of ultraviolet cutoff. Let us stress
that this regularization is gauge invariant.

The action can be rewritten in the following form (up to nonessential surface
terms):
 \disn{5.4.11}{
S=a^2\sum_{x^{\p}} \int dx^+ \left\{ \sum_i 4LF_{+-(0)}^{ii}
\dd_+v^i+\right. \hfill\no\hfill
\left. +\frac{i}{a^2}\sum_{i,j}\sum_k{\sum_n}'
{M^{ij}_{nk}}^+\dd_+M^{ij}_{nk}+2L \sum_i
A_{+(0)}^{ii}Q_{(0)}^{ii}-\tilde {\cal H}(x) \right\},
\nom}
where the $\sum'_n $ means that the sum is cut off by the condition
(\ref{5.15a}), and
the $\tilde{\cal H}$ is obtained from the ${\cal H}$ via
the substitution of the expression
 \disn{5.n91}{
F_{+-}^{ij}=\ls F_{+-}^{ij}-\de^{ij}F_{+-(0)}^{ii}\rs+
\de^{ij}F_{+-(0)}^{ii},
\nom}
where the $F_{+-}^{ij}-\de^{ij}F_{+-(0)}^{ii}$ are to be
written in terms of
$M^{ij}_{nk}$, ${M^{ij}_{nk}}^+$, $v^i$ by solving the constraints
(\ref{5.n8}) and using eq.~(\ref{5.4.10}).
The $F_{+-(0)}^{ii}$ remain independent. The $G_{12}^{ij}$ are also to be
expressed in terms of $M^{ij}_{nk}$, ${M^{ij}_{nk}}^+$, $v^i$
via the eqns.~(\ref{5.n1}), (\ref{5.4.10}).

We have the following set of canonically conjugated pairs of independent
variables:
 \disn{5.4.13}{
\left \{ v^i,\qquad \Pi^i=4La^2 F_{+-(0)}^{ii}\right \}, \no
\left \{ M^{ij}_{nk},\qquad i{M^{ij}_{nk}}^+ \right \}.
\nom}
In  quantum theory these variables become operators which satisfy
usual canonical commutation relations:
 \disn{5.n10}{
[v^i(x),\Pi^j(x')]_{x^+=0}=i\de^{ij}\de_{x^\p,x'^\p},\no
[M_{nk}^{ij}(x),{M_{n'k'}^{i'j'}}^+(x')]_{x^+=0}=
\de^{ii'}\de^{jj'}\de_{nn'}\de_{kk'}\de_{x^\p,x'^\p};
\nom}
the other commutators being equal to zero.

Let us remark that the condition (\ref{5.n5})
does not fix the gauge completely.
In particular, discrete group of gauge transformations,
depending on the $x^-$, of the form
 \disn{5.n10.1}{
 U_n^{ij}(x)=\de^{ij}\exp\left\{ i\frac{\pi}{L}n^i(x^{\p})x^-\right\},
 \nom}
where $n^i(x^{\p})$ are integers,
remains, and, of course, transformations, not depending on the $x^-$,
are allowed. Under the
transformations (\ref{5.n10.1}) canonical variables change as follows:
 \disn{5.n10.2}{
 v^i(x) \longrightarrow v^i(x)-\frac{\pi}{gL}\;n^i(x^{\p}),\qquad
 \Pi^i \longrightarrow \Pi^i,\no
 M_{nk}^{ij}(x^{\p}) \longrightarrow M_{n'k}^{ij}(x^{\p}),\quad
 n'=n+ n^i(x^{\p}) - n^j(x^{\p}- e_k).
 \nom}

Let us write the expression for quantum operators
$Q_{(0)}^{ii}(x^{\p},x^+)$,
which define the physical subspace of states.  We fix the order of the
operators in such a way as to relate with classical expression
$G_{\mu\nu}^{\;+}G^{\,\mu\nu}$
a quantum one of the form:
 \disn{5.n10.2a}{
 \frac{1}{2}\ls G_{\mu\nu}^{\;+}G^{\,\mu\nu} + G^{\,\mu\nu}G_{\mu\nu}^{\;+}\rs.
 \nom}
We remark that other choices of the ordering do not admit reasonable
vacuum solution.
Then the  operators $Q_{(0)}^{ii}(x^{\p},x^+)$
have the following form in terms of
canonical variables:
 \disn{5.4.16}{
2LQ^{ii}_{(0)}(x^\p,x^+)
=-\frac{g}{4a^2}
\sum_j\sum_k{\sum_n}'
\lks \e \ls p_n+gv^j(x+e_k)-gv^i(x)\rs\times\right.\hfill\no\times
\ls
{M^{ji}_{nk}}^+(x+e_k)M^{ji}_{nk}(x+e_k)+
M^{ji}_{nk}(x+e_k){M^{ji}_{nk}}^+(x+e_k)\rs-\no\hfill\left.
-\e\ls p_n+gv^i(x)-gv^j(x-e_k)\rs\ls
{M^{ij}_{nk}}^+(x)M^{ij}_{nk}(x)+
M^{ij}_{nk}(x){M^{ij}_{nk}}^+(x)\rs\rks,
\nom}
where
 \disn{5.n11}{
\e(p) = \cases{
$1$, & $p > 0$ \cr
$-1$, & $p < 0$ \cr}.
\nom}

One can easily construct canonical operator of translations in the $x^-$:
 \disn{5.4.14}{
 P_-^{can}=\frac{1}{2}\sum_{x^{\p}} \sum_{i,j}\sum_k{\sum_n}'
 p_n \e \ls p_n+gv^i(x)-gv^j(x-e_k) \rs\times\hfill\no\hfill\times\ls
 {M_{nk}^{ij}}^+(x)M_{nk}^{ij}(x)+M_{nk}^{ij}(x){M_{nk}^{ij}}^+(x)\rs.
 \nom}

This expression differs from the physical gauge invariant momentum
operator $P_-$ by a term proportional to the constraint. The operator $P_-$ is
 \disn{5.4.15}{
P_-=\frac{a^2}{2}\sum_{x^{\p}} \sum_k\; \int\limits_{-L}^L dx^-\, {\rm Tr}
\ls G^+_{-k}G_{-k}+G_{-k}G^+_{-k}\rs=
P_-^{can.} + 4La^2  \sum_{x^{\p}} \sum_i
v^iQ^{ii}_{(0)}=\hskip -3pt \no
=\frac{1}{2}\sum_{x^{\p}} \sum_{i,j}\sum_k{\sum_n}'
\left| p_n+gv^i(x)-gv^j(x-e_k) \right|
\times\hfill\no\hfill\times
\ls {M^{ij}_{nk}}^+(x)M^{ij}_{nk}(x)+M^{ij}_{nk}(x){M^{ij}_{nk}}^+(x)\rs
=\no=
\sum_{x^{\p}} \sum_{i,j}\sum_k{\sum_n}'
\left| p_n+gv^i(x)-gv^j(x-e_k) \right|\ls
{M^{ij}_{nk}}^+(x)M^{ij}_{nk}(x)+\frac{1}{2}\rs.
\nom}

Let us choose a representation of the state space, in which the variables
$v^i(x)$ are the multiplication operators. The states  are described in this
representation by normalizable functionals $F[v]$ of classical functions $v^i(x)$
(in fact by functions, depending on the values of the $v^i$ in different points
$x^\p$ due to the discreteness of these $x^\p$). One can define
full space of states as a direct product of the Fock space, in which the
${M^{ij}_{nk}}^+(x)$ and $M^{ij}_{nk}(x)$
play the role of creation and annihilation operators, and
the space of functionals $F[v]$.
Let us call $M$-vacuum the states of the form
$|0\ra\cdot F[v]$, where the $|0\ra$ satisfies the condition
 \disn{5.n12}{
 M^{ij}_{nk}(x)|0\ra=0,
 \nom}
and the $F$ is some functional. Arbitrary state can be represented in the form
of linear combination of vectors $|\{m\};F\ra$ of the form
 \disn{5.4.17}{
 \prod_{x^{\p}}\prod_{i,j}\prod_k{\prod_n}'
 \ls {M^{ij}_{nk}}^+(x)\rs^{m^{ij}_{nk}(x)}|0\ra\cdot F[v]
 \nom}
with different nonnegative integer functions  $m^{ij}_{nk}(x^\p)$
and functionals $F$. One can
define orthonormalized set of such functionals if necessary.

One can see from (\ref{5.4.15}) that the state, corresponding to the
absolute minimum of the $P_-$  must satisfy the conditions (\ref{5.n12}),
i.e. to be a $M$-vacuum.
The value of the $P_-$ in this state can be written in the form
 \disn{5.4.25}{
\langle\left. 0;F\right|P_-\left|\;0;F\right.\ra=
\hfill\no\hfill=
\frac{1}{2}\int \prod_{x^{\p}}\prod_i dv^i(x)
\sum_{x^{\p}}\sum_{i,j}\sum_k{\sum_n}'
\left|p_n+g v^i(x)-g v^j(x-e_k)\right| |F[v]|^2.
\nom}
Remind that the ${\sum\limits_n}'$
denotes the sum in $n$ limited by the condition
(\ref{5.15a}).
If one uses this condition and shifts the index $n$ in
these sums by integer part of the quantities
$(gL(v^i(x)-v^j(x-e_k))/\pi)$, one sees that the
dependence on the $v^i$ cancels in sums over $n$
and that the expression (\ref{5.4.25})
does not depend on the $F[v]$ if it is normalized to unity.
Thus the momentum
$P_-$ has the minimum in all $M$-vacua. One can make the value of the $P_-$ in
these vacua equal to zero by subtracting corresponding constant from the
operator $P_-$.

Let us show that $M$-vacua are the physical states,
i.~e. satisfy the condition
(\ref{5.4.9}).
Indeed, in $M$-vacua this condition looks as follows:
 \disn{5.4.24}{
\sum_j\sum_k{\sum_n}'\lks
\e\ls p_n+g v^j(x+e_k)-g v^i(x)\rs-
\right.\hfill\no\hfill-\left.
\e\ls p_n+g v^i(x)-g v^j(x-e_k)\rs\rks F[v]=0
\nom}
and is satisfied for any $F[v]$, because for every link in the sum
(\ref{5.4.24}) the numbers of positive and negative values of
the $\e$-functions are equal.
For  arbitrary  basic states (\ref{5.4.17})
analogous conditions have the following form:
 \disn{5.4.24a}{
\sum_j\sum_k{\sum_n}'\lks
\e\ls p_n+g v^j(x+e_k)-g v^i(x)\rs m^{ij}_{nk}(x+e_k)-\right.\hfill\no\hfill
-\left.\e\ls p_n+g v^i(x)-g v^j(x-e_k)\rs m^{ij}_{nk}(x)\rks F[v]=0.
\nom}

The eigen values $p_-$ of the operator $P_-$ can be found from the equation
 \disn{5.4.25a}{
\sum_{x^{\p}}\sum_{i,j}\sum_k{\sum_n}'
\left|p_n+g v^i(x)-g v^j(x-e_k)\right| m^{ij}_{nk}(x)F[v]=
p_- F[v]
\nom}
where the functional $F[v]$ is normalized.

To define  physical vacuum
state correctly one must consider not only states,
corresponding to the minimum of the operator $P_-$,
but also to the minimum of
the operator $P_+$. One can try to do this
via minimization  of the $P_+$
on the $M$-vacua, i.~e. on the set of states with $p_-=0$. The expression
$\langle 0|P_+|0\ra$,
where $|0\ra$ is the Fock space vacuum w.~r.~t. the $M_{nk}$, ${M_{nk}}^+$,
depends on the functions $v^i(x)$ (which enter, in particular, into
the "normal contractions"
of the operators $M_k$, $M_k^+$) and on the operators $\Pi^i$, canonically
conjugated to
the $v^i(x)$. The expectation value of this
expression is to be minimized on the set of
functionals $F[v]$.
The resulting functional $F[v]$ must decrease  in
the vicinity of those values of the $v^i(x)$, for which  the operator $D_-$
has zero eigen value, because
the Hamiltonian is singular at these values (it is seen, for example, from
the expansion (\ref{5.4.10})).

The vacuum state constructed in such a way
strongly differs from the usual vacuum in continuous space theory,
because for $M$-vacuum we get
zero expectation values of the operators $M_k$, but not of
the operators $(M_k - I)/ga = B_k + iA_k$,
related to usual gauge fields. Beside of this
the condition of the unitarity of the matrices $M_k$ in the continuum
limit (or equivalent condition of switching off the nonphysical fields $B_k$)
cannot be fulfilled in such a vacuum. This disagreement
with conventional theory is caused by the absence
of explicit Lorentz invariance in our formulation, that leads to
different quantum states, corresponding to the
minima of the operators $P_-$ and
$P_+$. It is not clear whether these states can be made coinciding at least in
the limit of continuous space. This requires further investigation.

Nevertheless the method of the quantization of gauge theories on the LF, described
here, can be useful for completely gauge-invariant formulation of some effective
models, based on analogous formalism (but
without complete gauge invariance due
to throwing out of all zero modes of fields and due to the absence of
gauge-invariant
regularization of ultraviolet divergencies in the $p_-$).
Such models are described, for example, in papers \cite{dalley1,dalley4},
where the ideas of papers \cite{pirner1,mack} were developed.

\section{Conclusion}

In the present paper complete gauge-invariant regularization of gauge field
theory in Hamiltonian approach on the LF is given with the help of
lattice formalism.
This regularization includes the limitation of the space  in the $x^-$,
$|x^-|\le L$, the periodic boundary conditions for fields in $x^-$ and
the two-dimensional lattice in transverse coordinates $x^{\p}$. Also we
introduced gauge-invariant regularization of ultraviolet divergencies in LF
momentum $p_-$ via the cutoff in modes of covariant derivative $D_-$.
Instead of
transverse components of gauge vector fields we used complex matrix
variables defined on lattice links. In these variables  the action can be
written in simple form, polynomial in the fields. In comparison
with  usual
field variables in continuous space, these matrix variables contain
a part, which reduces to some extra nonphysical fields in continuum
limit. The extra fields are to be switched off  via adding
to the Hamiltonian a
"mass" term for these fields with the "mass" fastly
rising in continuum limit.

We found that in our canonical LF formalism there are no 2nd class
constraints, connecting zero modes with other modes.
This allowed simple quantization of the
theory. The vacuum, defined  with respect
to the minimum
of the operator $P_-$, turned out to be different  from that of
continuous space
formulation, because the minimum of LF momentum $P_-$ does not
correspond to the minimum of the operator $P_+$.
This is due to the absence of
explicit Lorentz invariance in our regularization scheme. However it
remains open the question, whether a restoration of Lorentz invariance can be
achieved in the limit of continuous space.

Remarkably, our method can be applied to effective semiphenomenological,
"color dielectric" type, models where Lorentz invariance
can be partially restored via a modification of the LF Hamiltonian.

\end{document}